\documentclass[10pt,twocolumn,letterpaper]{article}

\usepackage{wacv}
\usepackage{times}
\usepackage{epsfig}
\usepackage{graphicx}
\usepackage{amsmath}
\usepackage{amssymb}
\usepackage{booktabs}

\usepackage{cite}
\usepackage{amsmath,amssymb,amsfonts}
\usepackage{algorithmic}
\usepackage{graphicx}
\usepackage{textcomp}
\usepackage{threeparttable}
\usepackage{xr}
\usepackage{multirow}
\usepackage{multicol}
\usepackage{stfloats}
\usepackage{sidecap}
\usepackage{subfig}
\usepackage{url}

\def\wacvPaperID{1475} 

\wacvfinalcopy 

\ifwacvfinal
\usepackage[breaklinks=true,bookmarks=false]{hyperref}
\else
\usepackage[pagebackref=true,breaklinks=true,colorlinks,bookmarks=false]{hyperref}
\fi

\newcommand*{\affaddr}[1]{#1}
\newcommand*{\affmark}[1][*]{\textsuperscript{#1}}

\makeatletter
\newcommand{\printfnsymbol}[1]{\textsuperscript{\@fnsymbol{#1}}}
\makeatother

\begin{document}

\title{UNet-2022: Exploring Dynamics in Non-isomorphic Architecture}

\author{Jiansen Guo\affmark[1]\thanks{First two authors contributed equally.}
\quad Hong-Yu Zhou\affmark[2]\printfnsymbol{1}\quad
Liansheng Wang\affmark[1]\quad Yizhou Yu\affmark[2]
\\
\affaddr{\affmark[1]School of Informatics, Xiamen University}\\
\affaddr{\affmark[2]Department of Computer Science, The University of Hong Kong}\\
\tt\small{jsguo@stu.xmu.edu.cn, whuzhouhongyu@gmail.com, lswang@xmu.edu.cn, yizhouy@acm.org}
}

\maketitle
\thispagestyle{empty}

\begin{abstract}
   Recent medical image segmentation models are mostly hybrid, which integrate self-attention and convolution layers into the non-isomorphic architecture. However, one potential drawback of these approaches is that they failed to provide an intuitive explanation of why this hybrid combination manner is beneficial, making it difficult for subsequent work to make improvements on top of them. To address this issue, we first analyze the differences between the weight allocation mechanisms of the self-attention and convolution. Based on this analysis, we propose to construct a parallel non-isomorphic block that takes the advantages of self-attention and convolution with simple parallelization. We name the resulting U-shape segmentation model as \emph{UNet-2022}. In experiments, UNet-2022 obviously outperforms its counterparts in a range segmentation tasks, including abdominal multi-organ segmentation, automatic cardiac diagnosis, neural structures segmentation, and skin lesion segmentation, sometimes surpassing the best performing baseline by 4\%. Specifically, UNet-2022 surpasses nnUNet, the most recognized segmentation model at present, by large margins. These phenomena indicate the potential of UNet-2022 to become the model of choice for medical image segmentation. Code is available at \url{https://bit.ly/3ggyD5G}.
\end{abstract}

\section{Introduction}
\label{sec:introduction}

\begin{figure*}[t]
\centerline{\includegraphics[width=1.0\textwidth]{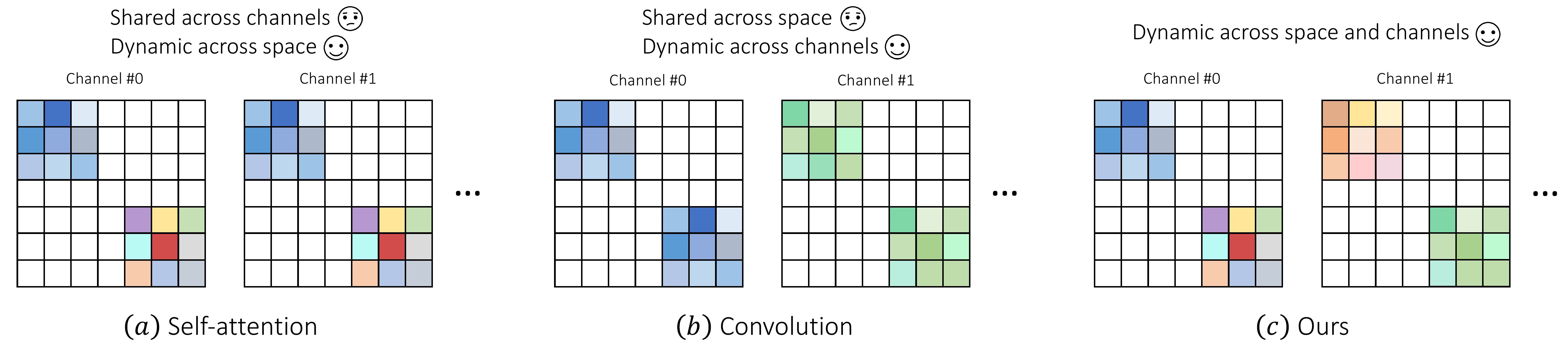}}
\caption{Illustration of the weight allocation mechanisms in self-attention (\textbf{a}), convolution (\textbf{b}), and our module (\textbf{c}). Different colors denote different weights. In self-attention, the weight matrix is dynamic across space but shared across different channels. In contrast, the weight matrix of convolution is shared across space but varies in channels. Our module integrates the advantages of self-attention and convolution by assigning dynamic weights to different positions and channels.}
\label{intro}
\end{figure*}

The last decade has witnessed the dominance of deep convolutional neural networks (DCNNs) in computer vision. However, in 2020, ViT~\cite{dosovitskiy2020image} showed that Transformers~\cite{vaswani2017attention}, which were initially developed for natural language processing (NLP), can perform equally well in vision tasks as DCNNs. Nonetheless, as pointed out by~\cite{dosovitskiy2020image,zhou2021convnets}, due to the lack of the locality inductive bias from DCNNs, Transformers do not generalize well when trained on insufficient amounts of data. To tackle this issue, Swin Transformer~\cite{liu2021swin} proposed to limit the self-attention computation to non-overlapping local windows. This design greatly reduces the computational cost and enables Swin Transformer outperform well-established DCNNs, such as ResNet~\cite{he2016deep}, in a wide range of vision tasks.

On the other hand, image segmentation has been among the fundamental tasks in medical image analysis. As the most widely adopted segmentation tools, UNet~\cite{ronneberger2015u} and most of its series~\cite{zhou2019unet++,li2018h,isensee2021nnu} were built upon DCNNs. With the prevalence of vision transformers in 2021, the medical imaging community started to incorporate the self-attention module into U-shape segmentation models for performance boosting~\cite{chen2021transunet,wang2021transbts,cao2021swin,xie2021cotr,chang2021transclaw,xu2021levit,zhou2021nnformer,peiris2021volumetric,huang2021missformer,liu2022phtrans}. The core behind these approaches is to construct non-isomorphic U-shape architecture by integrating self-attention with convolution. Although these methods achieved progress in different medical imaging tasks, most of them failed to provide an intuitive explanation for why this combination can be optimal. Accordingly, it is unclear how to better exploit the advantages of self-attention and convolution to build more optimal segmentation networks.

Let us briefly review the weight allocation mechanisms of self-attention and convolution, respectively. As is well-known, the key characteristic that lead to the success of Transformers is the self-attention mechanism~\cite{vaswani2017attention}. In vision transformers~\cite{dosovitskiy2020image,liu2021swin}, self-attention relates representations at different positions by employing a dynamic weight allocation mechanism. In practice, self-attention first computes the similarity between visual representations at different positions. Based on the resulting similarity matrix, dynamic weights are computed and assigned to representations at different spatial positions. Thus, as shown in Fig.~\ref{intro}(a) self-attention, different positions have different weights while all channels at the same position share the same weight. One potential problem of this design is that the weights assigned are not dynamic on the channel dimension, preventing self-attention from capturing the internal variances among different channels. 

On the other hand, DCNNs rely on extra learnable convolution kernels to aggregate spatial representations. As shown in Fig.~\ref{intro}(b) convolution, the same set of convolution kernel weights are shared across different spatial positions while dynamic weights are assigned to different channels. As a result, compared to self-attention, convolution can better explore the potential of representations emerged in different channels but lack the ability to describe complex spatial patterns.

In the above, we analyze the difference between self-attention and convolution from a perspective of weight allocation. From this analysis, we see that these two strategies maintain distinct but complementary characteristics. Based on this insight, we introduce a non-isomorphic block to include self-attention and convolution as two parallel modules. The proposed block comprises a novel weight allocation mechanism, which introduces dynamic to both space and channel dimensions (cf Fig.~\ref{intro}). In practice, we find this embarrassingly simple combination performs surprisingly well, outperforming previous state-of-the-art medical segmentation models by large margins in various segmentation tasks. Moreover, to reduce the risk of overfitting, we use depth-wise convolution (DWConv) for decreasing the number of weight parameters, which we empirically found performs slightly better than the naive convolution. To summarize, our contributions are as follows:
\begin{itemize}
    \item We provide an intuitive explanation for why self-attention and convolution can be complementary to each other. The core difference lies in the dynamic feature of weight allocation mechanism. Self-attention addresses the importance of spatial dynamic but ignores the channel dynamic. In contrast, convolution assigns dynamic weights to different channels instead of spatial positions.
    \item We propose a new weight allocation mechanism, introducing dynamic weights to both space and channel dimensions. The implementation of weight allocation mechanism is frustratingly simple, which comprises parallel independent self-attention and convolution modules. The resulting non-isomorphic block assigns dynamic weights to different spatial positions and channels, making it capable of capturing complex patterns emerged in feature maps.
    \item The resulting UNet-2022 obviously outperforms nnUNet, currently the best generic medical image segmentation model, in a range of medical image segmentation tasks, including abdominal multi-organ segmentation, automatic cardiac diagnosis, neural structures segmentation, and skin lesion segmentation. For instance, UNet-2022 surpasses nnUNet by nearly 4\% with a much smaller input size on multi-organ segmentation.
\end{itemize}

\section{Related work}
\label{sec:relatedwork}
\noindent \textbf{Vision Transformer.} Transformer was proposed in the NLP domain~\cite{vaswani2017attention} and maintains the ability of modeling long-range dependencies between elements of a sequence which can make up for the deficiencies of CNN in capturing the global information. ViT \cite{dosovitskiy2020image} first time applied Transformer to the vision tasks and achieved impressive performance that is comparable to or even better than the traditional DCNNs. However, ViT requires more data during the training stage due to the lack of locality inductive bias. To address this issue, a number of Transformer-based models were developed. On the basis of the ViT, DeiT~\cite{touvron2021training} introduced stronger data augmentation to regularize vision transformers. Besides, DeiT employed an idea of knowledge distillation to help train vision transformers, where the teacher network could help the student network to incorporate the locality inductive bias. Swin Transformer~\cite{liu2021swin} built a hierarchical vision Transformer that introduces the window-based self-attention mechanism to enhance the ability to capture the locality of features and reduces the computation complexity of the self-attention. The local self-attention is more suitable for the dense prediction tasks such as semantic segmentation. However, most of these work failed to notice the intrinsic drawback of the weight allocation mechanism used in the self-attention, making them often less competitive than hybrid models. \\

\noindent \textbf{Hybrid medical image segmentation models.} TransUNet~\cite{chen2021transunet}, TransClaw UNet~\cite{chang2021transclaw}, and LeViT-UNet \cite{xu2021levit} inserted self-attention layers between the encoder and decoder of DCNNs to take advantage of capturing long-range dependencies among a number of feature channels. Swin UNet~\cite{cao2021swin} replaced the convolutional blocks with the Swin Transformer blocks~\cite{liu2021swin} and built a U-shape segmentation model. DS-TransUNet~\cite{lin2022ds} extended Swin UNet by introducing a fusion module for modeling long-range dependencies between features of different scales. Similar to Swin UNet, MISSFormer~\cite{huang2021missformer} built a hierarchical U-shape Transformer network that bridges all stages from the encoder to the decoder by mixing multi-scale information obtained by the hierarchical Transformer blocks. UNETR~\cite{hatamizadeh2022unetr} adopted ViT~\cite{dosovitskiy2020image} as the encoder network. In UNETR, feature maps from different layers of ViT with different resolutions are collected and sent to the convolutional decoder to capture the multi-scale information. nnFormer~\cite{zhou2021nnformer} utilized both local and global volume-based self-attention to build feature pyramids. MedT~\cite{valanarasu2021medical} proposed a gated axial attention layer which introduces a summational control mechanism in the self-attention. However, one problem of these hybrid segmentation models is that they did not provide an intuitive explanation of why the combination of self-attention and convolution can be beneficial. As a result, it is still unclear how to build an optimal combination of self-attention and convolution.\\

\begin{figure*}
    \centering
    \subfloat[Overall architecture]{\includegraphics[height=0.3\textwidth]{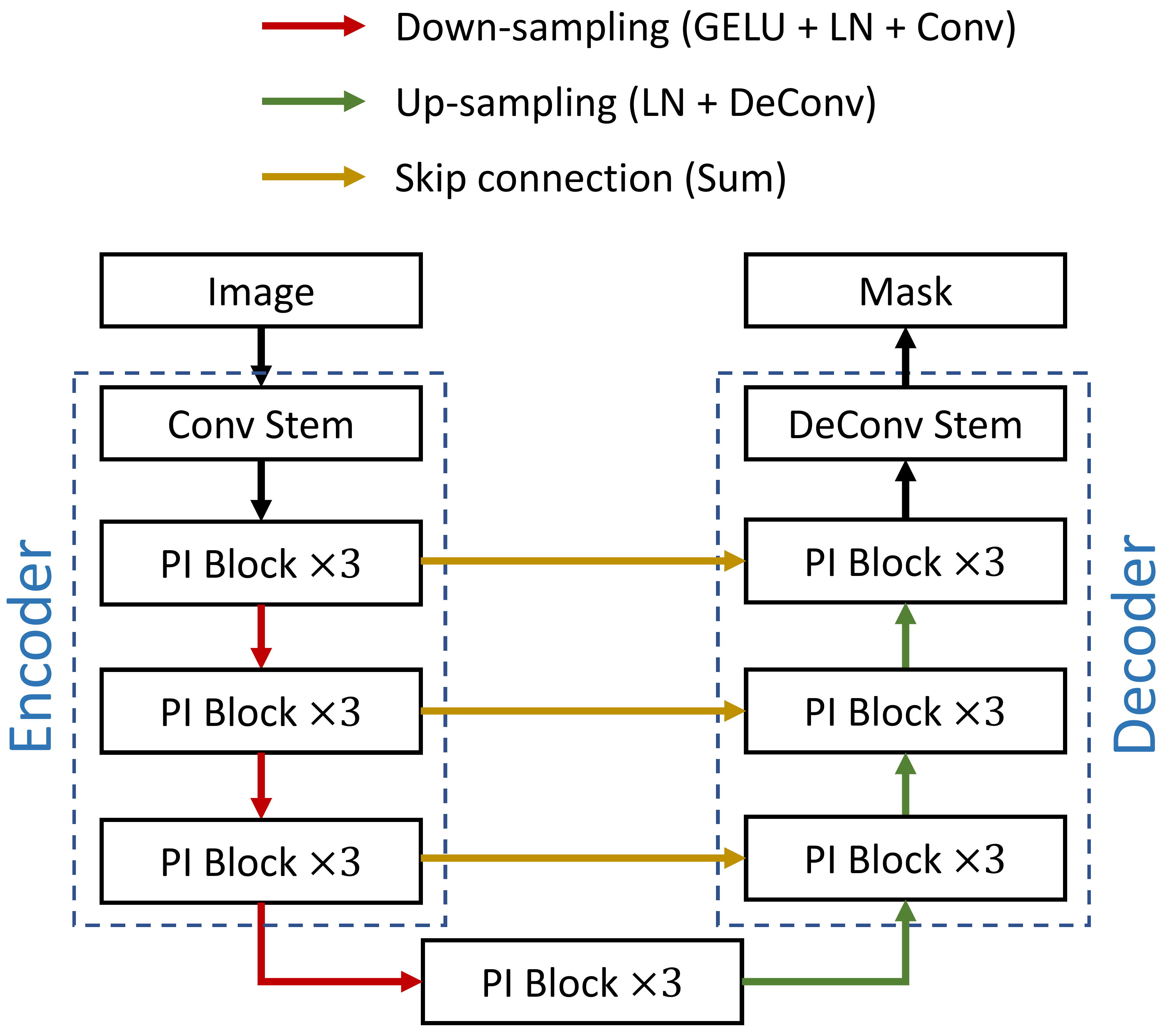}\label{m1}}\ \ 
    \subfloat[Parallel non-isomorphic block]{\includegraphics[height=0.3\textwidth]{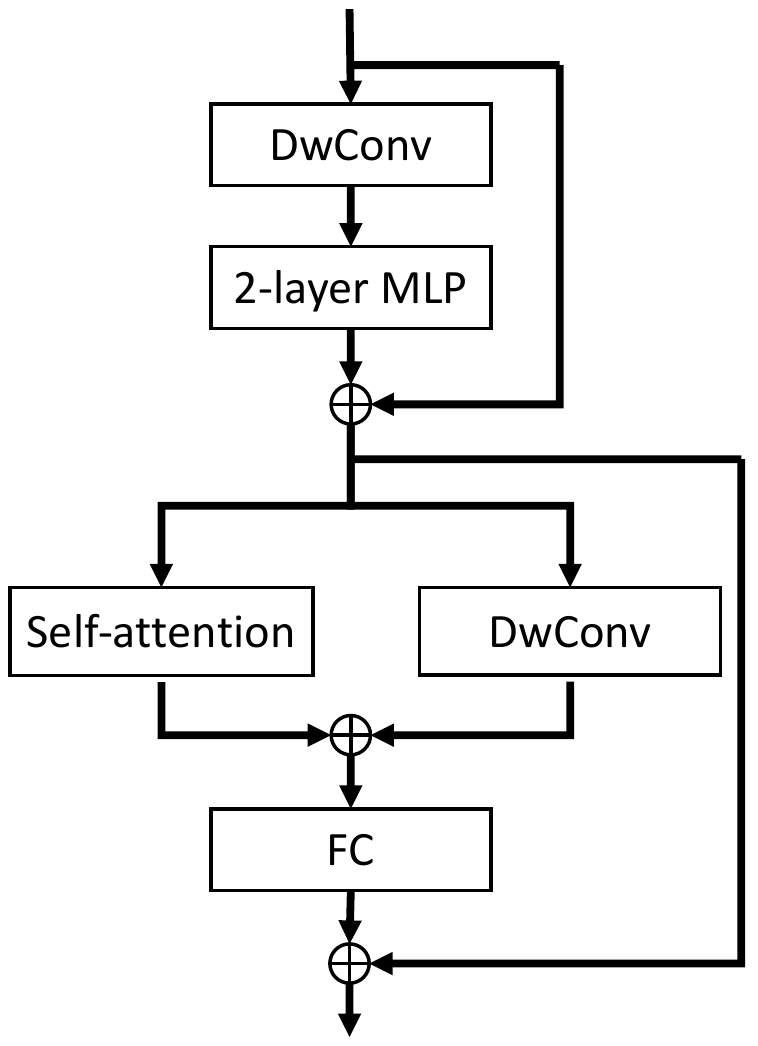}\label{m2}}\ \ 
    \subfloat[Convolution stem]{\includegraphics[height=0.3\textwidth]{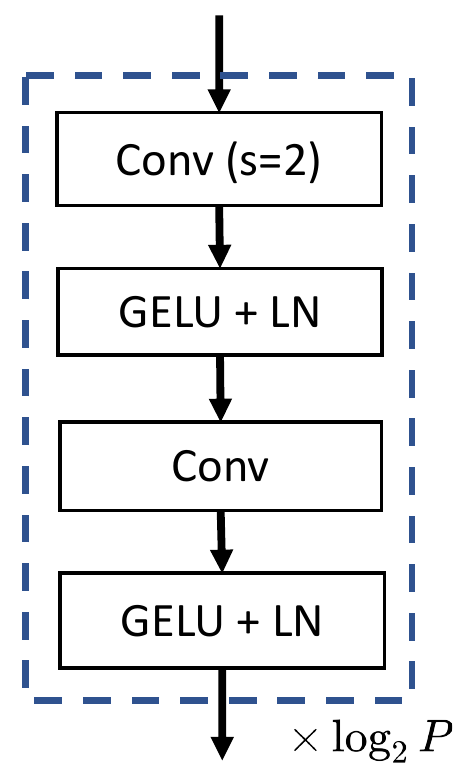}\label{m3}}\qquad
    \subfloat[De-convolution stem]{\includegraphics[height=0.3\textwidth]{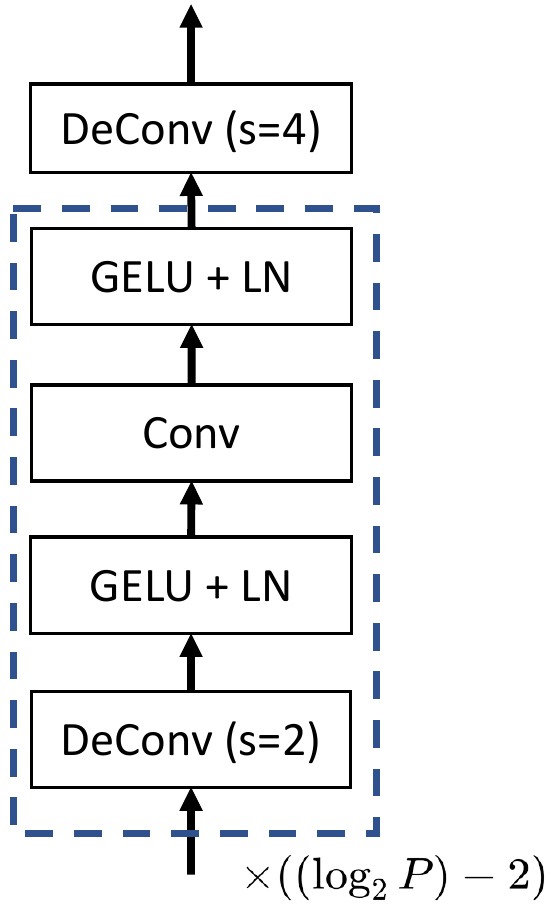}\label{m4}}
    \caption{Illustrations of UNet-2022. In part (\textbf{a}), we present the overall architecture of UNet-2022. Part (\textbf{b}) displays the internal structures of the proposed parallel non-isomorphic (PI) block. In parts (\textbf{c}) and (\textbf{d}), we show the detailed structures of the convolution (Conv) and de-convolution (DeConv) stems. \textbf{DwConv}, \textbf{MLP}, \textbf{FC}, and \textbf{LN} stand for the depth-wise convolution, multi-layer perceptron, fully-connection layer, and layer normalization. $P$ is a hyper-parameter, which varies based on the input resolution.}
    \label{method}
\end{figure*}

\section{Methodology}
We illustrate the detailed architecture of UNet-2022 in Fig.~\ref{method}. As shown in Fig.~\ref{m1}, the encoder of UNet-2022 consists of one convolution stem and three stages, where each stage involves three parallel non-isomorphic (PI) blocks. Symmetrically, the decoder also comprises three stages and one de-convolution stem. At each down-sampling/up-sampling step, we increase/decrease the number of channels and decrease/increase the spatial resolution of feature maps accordingly. Skip connections are used to bridge the gap between low-level details and high-level semantics.

Fig.~\ref{m2} describes the internal structural details of the parallel non-isomorphic block. We use depth-wise convolution (DwConv) to reduce the number of parameters. Self-attention and depth-wise convolution are parallelized to explore dynamics in spatial and channel dimensions, respectively, whose outputs are then added up and passed to a fully-connected (FC) layer. In the convolution stem, we stack multiple convolution layers to extract high-resolution feature maps, inspired by~\cite{zhou2021nnformer}. Similar operations are also applied in the de-convolution stem, where multiple de-convolution layers are stacked to produce the final segmentation mask.

\subsection{Weight allocation in convolution and self-attention}
Either self-attention or convolution can be regarded as a per-dimension weight allocation function. Suppose the feature map ${\mathcal{F}}^l\in \mathbb{R}^{H\times W\times C}$ contains $N$ overlapping local patches, where $N=H\times W$ denotes the number of patches in ${\mathcal{F}}^l$. $H$ and $W$ are the height and width of the 2D feature map, and $C$ is the number of channels. We use $\mathcal{V}^l_t\in \mathbb{R}^{K\times K\times C}$ to denote the local representation patch centered at the $t$-th position, $t\in\{0,1,...,N-1\}$. Accordingly, we use $\mathbf{W}^{\text{sa}}_t\in \mathbb{R}^{K\times K \times C}$ to denote the assigned weight tensor in self-attention. Specifically, we investigate the weight allocation mechanism of depth-wise convolution, which is used to replace convolution in parallel non-isomorphic blocks. $\mathbf{W}^{\text{cv}}_{t,c}\in \mathbb{R}^{K\times K}$ denotes the weight kernel for the $c$-th channel in depth-wise convolution, $c\in \{0,1,...,C-1\}$. We ignore the layer index $l$ in $\mathbf{W}^{\text{sa}}_t$ and $\mathbf{W}^{\text{cv}}_{t,c}$ for presentation convenience.

For self-attention, $\mathbf{W}_{\text{sa}}$ is computed following a spatially dynamic manner, which can be summarized as:
\begin{align}
    \begin{split}
        \mathbf{W}^{\text{sa}}_t [i,j] = \frac{\langle \mathcal{V}_t^l[i,j],  \mathcal{V}_t^l[i,j] \rangle}{\sum_{p}^K \sum_{q}^K \langle \mathcal{V}_t^l[i,j], \mathcal{V}_t^l[p,q] \rangle}.
    \end{split}
    \label{eq:1}
\end{align}
$\{i, j, p, q\}$ denote the spatial indices. $\langle\cdot,\cdot \rangle$ stands for the inner product (of two vectors). From Eq.~\ref{eq:1}, we see that $\mathbf{W}_t^{\text{sa}}$ varies in the spatial dimension but remains unchanged in the channel dimension.

For depth-wise convolution, $\mathbf{W}^{\text{cv}}_{t,c}$ is a learnable parameter, which is shared across $N$ different spatial positions in the $c$-th channel. Different from $\mathbf{W}^{\text{sa}}_t$, $\mathbf{W}^{\text{cv}}_{t,c}$ varies across the channel dimension, enabling it to capture inter-channel variations.

\subsection{Parallel non-isomorphic block}
As aforementioned, self-attention and convolution emphasize dynamics in different dimensions, making them complementary to each other. Inspired by this finding, we propose to integrate their advantages in a non-isomorphic block via straightforward parallelization on self-attention and convolution.

Suppose $\mathcal{F}^l\in \mathbb{R}^{H\times W\times C}$ denotes the input feature map to a parallel non-isomorphic block. As shown in Fig.~\ref{m2}, we first pass the input feature map $\mathcal{F}^l$ to a DwConv layer, after which a 2-layer multi-layer perceptron (MLP) is appended. The internal structures of the MLP layer are as follows: LayerNorm (LN)-FC-GELU-FC. We also add a residual connection to the output of MLP:
\begin{align}
    \begin{split}
    \hat{\mathcal{F}}^l = \text{MLP}(\text{DwConv}(\mathcal{F}^l)) + \mathcal{F}^l.
    \end{split}
    \label{eq:2}
\end{align}
$\hat{\mathcal{F}}^l$ is then forwarded to two parallel layers,i.e., self-attention and DwConv, whose outputs are added up:
\begin{align}
    \begin{split}
        \Tilde{\mathcal{F}}^l = \text{SA}(\hat{\mathcal{F}}^l) + \text{DwConv}(\hat{\mathcal{F}}^l),
    \end{split}
\end{align}
where SA stands for the self-attention layer. Here we employed the window-based self-attention layer, proposed in~\cite{liu2021swin}, to improve the running efficiency. The kernel size of DwConv is 7. Finally, $\Tilde{\mathcal{F}}^l$ is fed to the last FC layer and another residual connection is added:
\begin{align}
    \begin{split}
        \overline{\mathcal{F}}^l = \text{FC}(\Tilde{\mathcal{F}}^l) + \hat{\mathcal{F}}^l.
    \end{split}
\end{align}
$\overline{\mathcal{F}}^l$ is the output of the PI block and will be passed to the following layer as the input feature map.

\subsection{Convolution stem}
Convolution stems have been shown to be more advantageous in providing high-resolution representations with rich details~\cite{zhou2021nnformer,liu2021swin}. As displayed in Fig.~\ref{m3}, the convolution stem consists of a number of stacked blocks (layers in the dashed box). Each block is composed of two convolution layers with stride 2 and stride 1, respectively. Each convolution layer is followed by one GELU and one layer normalization. Each time when the spatial resolution of the feature map is reduced by half, we double its channel number accordingly.

To adapt to images with high resolutions, we stack more blocks in the convolution stem in order to reduce the memory cost and improve the computational efficiency. Specifically, the number of blocks, i.e., $\log_{2}{P}$, depends on the patch size $P$ that we manually set. For instance, we set $P$ to 4 for small and medium input resolutions, such as $224\times 224$ and $320\times 320$. We increase $P$ to 8 when the input resolution is high, such as $512\times 512$. Compared to the patchify stem used in~\cite{liu2021swin,dosovitskiy2020image}, we empirically found our convolution stem could better capture the low-level information with equal receptive fields.

\subsection{De-convolution stem}
Similar to the convolution stem, the architecture of the de-convolution stem also varies based on the resolution of input images. For small ($224\times 224$) and medium ($320\times 320$) input sizes, the de-convolution stem only consists of one de-convolution layer, whose kernel size is 4 and stride is 4. As for the high-resolution inputs (i.e., $512\times 512$), we adopt a similar strategy as used in the convolution stem, which is increasing the number of up-sampling blocks (cf. Fig.~\ref{m4}). In practice, the de-convolution stem involves blocks with de-convolutional layers to achieve goal of up-sampling feature maps for producing the final mask predictions. Specifically, the number of the de-convolution blocks is $(\log_{2}{P})-2$, which is 0 for small ($P$=4) and medium ($P$=4) resolutions, and 1 for the high-resolution input ($P$=8), respectively.

\section{Experiments}
\label{sec:experiment}
We conduct experiments on various medical image segmentation tasks, including abdominal multi-organ segmentation (Synapse), automatic cardiac diagnosis (ACDC), neural structures segmentation (EM), skin lesion segmentation (ISIC-2016 and PH2). On Synapse and ACDC, we transform CT/MRI volume data into slices and treat these slices as the training data. Experimental results show that our UNet-2022 achieves obvious performance gains over previous DCNN- and Transformer-based segmentation models.

\subsection{Dataset}
\noindent \textbf{Multi-organ CT segmentation (Synapse).} Synapse\footnote{\url{https://www.synapse.org/##!Synapse:syn3193805/wiki/217789}} consists of 30 abdominal CT scans, where 13 organs were annotations. After pre-processing, we extract 3,779 slices from all CT cases. Following instructions from~\cite{chen2021transunet}, we split the whole dataset into training (18 scans, 2,211 slices) and test (12 scans, 1,568 slices) sets. The Dice score (DSC) and 95$\%$ Hausdorff Distance (HD95) are reported on 8 abdominal organs, which are the aorta, gallbladder, spleen, left kidney, right kidney, liver, pancreas, and stomach.\\

\noindent \textbf{Automated cardiac diagnosis (ACDC).} ACDC~\cite{bernard2018deep} involves 100 samples, with the cavity of the right ventricle (RV), the myocardium of the left ventricle (MYO), and the cavity of the left ventricle (LV) to be segmented. After pre-processing, we obtain 1,902 slices. The dataset is split into 70 samples for training (1,290 slices), 10 samples for validation (196 slices), and 20 samples for testing (416 slices). DSC is used as the major evaluation metric. \\

\noindent \textbf{Neural structures segmentation (EM).} EM (Electron Microscopy) dataset~\cite{cardona2010integrated} contains 30 images and the size of each image is $512\times 512$. The whole dataset is split into 24 samples for training, 3 samples for validation, and 3 samples for test. Intersection over Union (IoU) is used as the evaluation metric. Similar to~\cite{zhou2019unet++}, we calculate multiple IoUs at thresholds ranging 0.5 from 0.95 with a step size of 0.05.\\

\noindent \textbf{Skin lesion segmentation (ISIC-2016 and PH2).} The training dataset comes from the International Skin Imaging Collaboration at year 2016 (ISIC-2016), which contains 900 samples with lesion segmentations from dermoscopic images. Following~\cite{lee2020structure,wang2021boundary}, we construct the test set using images from PH2~\cite{barata2013two}. DSC and IoU are used as evaluation metrics. \\

\subsection{Implementation details}
All experiments are implemented on a single NVIDIA 2080ti GPU with 11GB memory. We utilize both cross-entropy loss and dice loss and add them up like the Equation~\ref{loss} where the $\lambda_1$ and $\lambda_2$ are 1.2 and 0.8 for all datasets.
\begin{align}
    \begin{split}
        \mathcal{L}=\lambda_1\mathcal{L}_{\text{DSC}} + \lambda_2\mathcal{L}_{\text{CE}}.
    \end{split}
    \label{loss}
\end{align}
On Synapse, we make experiments on two different input resolutions, i.e., $224\times 224$ and $320\times 320$. For ACDC, we adopt the small input size but increase the embedding dimension to 192. For skin lesion and neural structures segmentation tasks, we fix the crop size and embedding dimension to $512\times 512$ and 96, respectively. The default optimizer is Adam~\cite{kingma2014adam}. We fix the learning rate during the training stage. Please refer to the supplementary material for more network configuration details.\\

\noindent \textbf{Deep supervision.} We add deep supervision to networks when training segmentation models on Synapse, ISIC-2016, and EM datasets. In practice, we collect the decoder's outputs in three different stages (with different resolutions) and pass them to independent de-convolution stems. Then, we calculate the loss function (cf. Eq.~\ref{loss}) on the predicted masks (with three different resolutions). The concrete procedure can be summarized as follows: 
\begin{align}
    \begin{split}
        \mathcal{L}_{all}=\alpha_1\mathcal{L}_{\{H,W \}} + \alpha_2\mathcal{L}_{\{\frac{H}{2},\frac{W}{2}\}} +
        \alpha_3\mathcal{L}_{\{\frac{H}{4},\frac{W}{4}\}},
    \end{split}
    \label{loss}
\end{align}
where $\alpha_1=\frac{1}{2}$, $\alpha_2=\frac{1}{4}$, $\alpha_3=\frac{1}{8}$. $\mathcal{L}_{H,W}$ denotes we calculate the loss function on the predicted mask whose resolution is $H\times W$. The same situation applies to $\mathcal{L}_{\{\frac{H}{2},\frac{W}{2}\}}$ and $\mathcal{L}_{\{\frac{H}{4},\frac{W}{4}\}}$.\\

\noindent \textbf{Augmentation strategies.} The patches sent to the network are first randomly cropped from the complete images. Then, augmentations such as rotation, scaling, gaussian noise, gaussian blur, brightness and contrast adjust, simulation of low resolution, gamma augmentation, and mirroring are applied in the given order during the training process. Note that we apply aggressive rotation augmentations to each image where the rotation angle will be randomly chosen from $[0,15]$. \\

\noindent \textbf{Inference details.} During the inference stage, UNet-2022 makes predictions following a sliding window manner. On Synapse, we set the step size of each sliding window to \textsc{0.3$\times$crop size} and \textsc{0.2$\times$crop size} for $224\times 224$ and $320\times 320$ input sizes, respectively. A smaller step size means that more overlapped patches participate in the voting of the mask prediction, leading to better segmentation performance. On the rest three datasets, the crop size is close to the full image size. Thus, adjusting the step size will not have observable impacts on the segmentation performance. For patch voting, we employ the gaussian importance weighting strategy, which gives the pixel at the center more weight in the softmax aggregation process.\\
\begin{table*}[t]

	\resizebox{\textwidth}{!}{
	\begin{tabular}{|l|c|cc|c|c|c|c|c|c|c|c|} 
	\hline
	    \multirow{2}{*}{Methods} &\multirow{2}{*}{Size} & \multicolumn{2}{c|}{Average} & \multirow{2}{*}{Aorta} & \multirow{2}{*}{Gallbladder}  & \multirow{2}{*}{Kidney (Left)}  & \multirow{2}{*}{Kidney (Right)}  & \multirow{2}{*}{Liver}  & \multirow{2}{*}{Pancreas} & \multirow{2}{*}{Spleen}  & \multirow{2}{*}{Stomach} \\ \cline{3-4}
	    & & DSC $\uparrow$ & HD95 $\downarrow$  &&&&&&&&\\
		\hline
		\hline
		ViT~\cite{dosovitskiy2020image} + CUP~\cite{chen2021transunet} & 224 & 67.86 &36.11 & 70.19 & 45.10 & 74.70 & 67.40 & 91.32 & 42.00 & 81.75 & 70.44\\
		R50-ViT~\cite{dosovitskiy2020image}  + CUP~\cite{chen2021transunet} & 224 &71.29 &32.87 & 73.73 & 55.13 & 75.80 & 72.20 & 91.51 & 45.99 & 81.99 & 73.95\\
        TransUNet \cite{chen2021transunet} & 224 &77.48  &31.69  &  87.23 & 63.16 & 81.87 & 77.02 & 94.08 & 55.86 & 85.08 & 75.62\\
        TransUNet \cite{chen2021transunet} & 512 &84.36  &-  & 90.68 & \textbf{71.99} & 86.04 & 83.71 & 95.54 & 73.96 & 88.80 & \underline{84.20}\\
        TransClaw UNet \cite{chang2021transclaw} & 224 & 78.09  &26.38  & 85.87 & 61.38 & 84.83 & 79.36 & 94.28 & 57.65 & 87.74 &73.55\\
        TransClaw UNet \cite{chang2021transclaw} & 512 &80.39 &- &90.00 &56.86 &83.27 &76.21 &95.06 &67.76 &91.16 &82.82\\
        SwinUNet \cite{cao2021swin} &224&79.13  &21.55  & 85.47 & 66.53 &83.28 &79.61 & 94.29 & 56.58 & 90.66 & 76.60\\
        SwinUNet{$^\bigtriangledown$} \cite{cao2021swin}&384
        & 81.12 & - &87.07&70.53&84.64&82.87&94.72&63.73&90.14&75.29\\
        LeViT-UNet-384s	\cite{xu2021levit} &224&78.53&16.84 &87.33&62.23	&84.61	&80.25	&93.11	&59.07	&88.86	&72.76\\
        MT-UNet \cite{wang2021mixed} &224&78.59 &26.59 &87.92 &64.99 &81.47 &77.29 &93.06 &59.46 &87.75 &76.81\\ 
        MISSFormer \cite{huang2021missformer} &224&81.96  &18.20  & 86.99 & 68.65 & 85.21 & 82.00 & 94.41 & 65.67 & \underline{91.92} & 80.81\\
        nnUNet \cite{isensee2021nnu} &512 &82.36 &24.74 &90.96 &65.57 &81.92 &78.36 &95.96 &69.36 &91.12 &85.60\\
        \hline
        UNet-2022 & 224&\underline{84.98} &\underline{16.70} &\textbf{92.10} &\underline{69.63} &\underline{88.40} &\underline{83.93} &\underline{96.02} &\underline{75.50} &90.40 &83.86\\
        \hline
        UNet-2022 & 320 &\textbf{86.46} &\textbf{11.34} &\underline{91.96} &69.40 &\textbf{89.26} &\textbf{85.58} &\textbf{96.34} &\textbf{75.66} &\textbf{94.22} &\textbf{89.29}\\
        \hline
	\end{tabular}
	}
\caption{Comparisons with 2D DCNN-based and hybrid segmentation models on multi-organ segmentation (Synapse). The evaluation metrics are DSC (\%) and HD95 (mm). The best results are bolded while the second best are underlined. We also investigate the impact of the input resolution. Besides $224\times 224$, TransUNet, TransClaw UNet, and nnUNet use a size of $512\times 512$. SwinUNet uses a size of $384\times 384$ while our UNet-2022 uses a size of $320\times 320$. $\uparrow$ means the higher the better and $\downarrow$ represents the opposite.}
\label{synapse}
\end{table*}

\begin{table}[t]
    \begin{center}
    \resizebox{\columnwidth}{!}{
	\begin{tabular}{|l|c|ccc|} 
	\hline
		Methods & Ave. DSC $\uparrow$ & RV & Myo & LV \\
		\hline
		\hline
        ViT \cite{dosovitskiy2020image} + CUP \cite{chen2021transunet}	&81.45	&81.46	&70.71	&92.18\\
        R50-VIT \cite{dosovitskiy2020image} + CUP \cite{chen2021transunet}	&87.57	&86.07	&81.88	&94.75\\
        TransUNet \cite{chen2021transunet}&89.71	&88.86	&84.54	&95.73\\
        SwinUNet \cite{cao2021swin}&90.00	&88.55	&85.62	&95.83\\
        LeViT-UNet-384s	\cite{xu2021levit}& 90.32 & 89.55 & 87.64 & 93.76\\
        MISSFormer \cite{huang2021missformer}&90.86 &89.55 &88.04 &94.99\\
        MT-UNet \cite{wang2021mixed} &90.43 &86.64 &89.04 &95.62\\
        nnUNet \cite{isensee2021nnu} &\underline{92.32} &\underline{90.39} &\underline{90.53} &\underline{96.05}\\
        \hline
        UNet-2022 &\textbf{92.83} &\textbf{91.04}	&\textbf{90.97}	&\textbf{96.49}\\
        \hline
	\end{tabular}
	}
    \end{center}
\caption{Comparisons with 2D DCNN-based and hybrid segmentation models on automatic cardiac diagnosis (ACDC). The evaluation metric is DSC (\%). The best results are bolded while the second best are underlined. The default input size is $224 \times 224$ for all approaches.}
\label{ACDC}
\end{table}

\subsection{Comparisons on abdominal multi-organ segmentation}
Table~\ref{synapse} presents the segmentation performance on 8 organs on Synapse. When the input size is $224\times 224$, we see that MISSFormer~\cite{huang2021missformer} achieves the highest average DSC while LeViT-UNet-384s~\cite{xu2021levit} produces the lowest average HD95 among all baselines. In comparison, our UNet-2022 outperforms MISSFormer by about 3\% in the average DSC. For specific organs, our UNet-2022 outperforms MISSFormer on 7 abdominal organs. These improvements include 5\% on aorta, 1\% on gallbladder, 3\% on left kidney, 2\% on right kidney, 2\% on liver, 10\% on pancreas, and 3\% on stomach. Despite average HD95 of our method is slightly better than that of LeViT-UNet-384s, our approach achieves a much higher average DSC, surpassing LeViT-UNet-384s by over 6\%.

On the other hand, as shown in previous work~\cite{chen2021transunet,chang2021transclaw,cao2021swin}, increasing the input resolution often brings observable improvements to the segmentation performance. Considering the trade-off between the running efficiency and task performance, we increase the input resolution to $320\times 320$ and observe large performance gains over $224\times 224$. Compared to TransUNet~\cite{chen2021transunet} that uses a much larger input size $512\times 512$, UNet-2022 ($320\times 320$) brings a 2-percent improvement to the average DSC. Similar situations also apply to TransClaw UNet ($512\times 512$) and SwinUNet ($384 \times 384$). Comparing UNet-2022 with nnUNet~\cite{isensee2021nnu}, our method achieves impressive progress under both DSC and HD95 metrics. UNet-2022 outperforms nnUNet by 4\% in the average DSC while dramatically reducing HD95 to 11.34mm (14mm reduction). 

\subsection{Comparisons on automated cardiac diagnosis}
In Table~\ref{ACDC}, we compare the segmentation performance of different models on ACDC. We fix the input resolution to $224\times 224$. Similar to the results displayed in Table~\ref{synapse}, MISSFormer achieves the highest average DSC among all hybrid segmentation models. However, compared to MT-UNet~\cite{wang2021mixed}, MISSFormer produces lower DSC on Myo and LV, implying the instability of it. In comparison, our UNet-2022 outperforms all hybrid segmentation baselines by large margins. Specifically, UNet-2022 surpasses MISSFormer by nearly 2\% in average.

Somewhat surprisingly, we find that nnUNet obviously outperforms MISSFormer by nearly 1.5\% on average while providing consistent performance gains on all three classes. Nonetheless, our UNet-2022 still outperforms nnUNet by about 0.5\% in average. Moreover, UNet-2022 achieves consistent improvements on all three individual classes, demonstrating the potential of UNet-2022 to replace nnUNet.

\begin{table*}[!htpb]

    \begin{center}
    \resizebox{0.8\textwidth}{!}{
	\begin{tabular}{|c|c|c|c|c|c|c|c|c|} 
	\hline
		Methods & UNet \cite{falk2019u} & Wide UNet \cite{zhou2019unet++}&UNet+ \cite{zhou2019unet++}&UNet++ \cite{zhou2019unet++}&nnUNet \cite{isensee2021nnu}&UNet-2022\\
		\hline
		\hline
        mIoU $\uparrow$&88.30 &88.37&88.89&89.33&\underline{90.55} &\textbf{91.05}\\
        \hline
	\end{tabular}
	}
	\end{center}
\caption{Comparisons with UNet series on neural structures segmentation (EM). The evaluation metric is the mean IoU (mIoU), where we calculate multiple IoUs at thresholds ranging 0.5 from 0.95 with a step size of 0.05. The best result is bolded while the second best is underlined. $\uparrow$ means the higher the better.}
\label{EM}
\end{table*}

\begin{figure}[t]
\centerline{\includegraphics[width=0.5\textwidth]{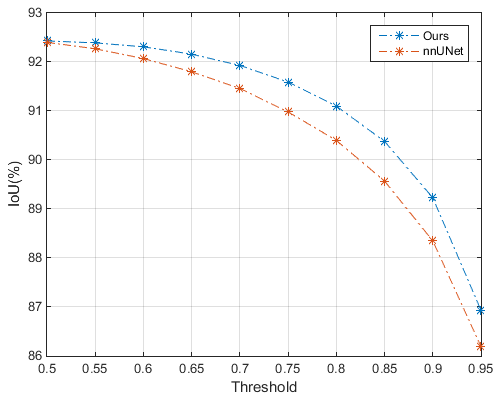}}
\caption{Comparisons of nnUNet and UNet-2022 on neural structures segmentation (EM). We present the computed IoU results at thresholds ranging from 0.5 to 0.95 with a step size of 0.05.}
\label{threshold}
\end{figure}

\begin{table}[!tpb]

    \begin{center} 
    \resizebox{0.3\textwidth}{!}{
	\begin{tabular}{|l|c|c|} 
	\hline
		Methods & DSC $\uparrow$& IoU $\uparrow$\\
		\hline
		\hline
        SSLS \cite{ahn2015automated} & 78.3 & 68.1   \\
        MSCA \cite{bi2016automated}	& 81.5 & 72.3 \\
        FCN \cite{long2015fully} &89.4 & 82.1 \\
        Bi $et~al$ \cite{bi2017dermoscopic}	& 90.6 & 83.9 \\
        nnUNet~\cite{isensee2021nnu} & 91.6 & 85.1\\
        Lee $et~al$ \cite{lee2020structure} & 91.8 & 84.3 \\
        BAT~\cite{wang2021boundary} & \underline{92.1} & \underline{85.8} \\
        \hline
        UNet-2022 &\textbf{93.6} &\textbf{88.4}\\
        \hline
	\end{tabular}
	}
	\end{center}
\caption{Comparisons with 2D DCNN-based and hybrid segmentation models on skin lesion segmentation. The evaluation metric is DSC (\%) and IoU(\%). The best results are bolded while the second best are underlined.}
\label{ISIC2016}
\end{table}

\subsection{Comparisons on neural structures segmentation}
In this task, we follow~\cite{zhou2019unet++} to calculate IoUs at thresholds ranging from 0.5 to 0.95 with a step size of 0.05. In Table~\ref{EM}, we present the segmentation performance of a range of UNet-like models. We see that nnUNet is the best performing baseline, outperforming the second best UNet++ by over 1\%. Thus, we thoroughly compare UNet-2022 against nnUNet at different thresholds of IoU in Fig.~\ref{threshold}. The default threshold is usually set as 0.5. As Fig.~\ref{threshold} displays, compared to nnUNet, UNet-2022 has obvious advantages in large thresholds. This phenomenon indicates that UNet-2012 is more advantageous in making high-confidence predictions.

\subsection{Comparisons on skin lesion segmentation}
Table~\ref{ISIC2016} presents the comparisons of DCNN-based and hybrid models on skin lesion segmentation. The best performing baseline is BAT~\cite{wang2021boundary}, which is a customized hybrid segmentation model that integrates the boundary-aware attention. BAT outperforms nnUNet by 0.5 and 0.7\% in DSC and IoU, respectively. 

Comparing UNet-2022 with BAT, we see that UNet-2022 achieves dramatic improvements in both DSC and IoU. For instance, UNet-2022 outperforms BAT by 2.6\% in IoU while BAT only surpasses nnUNet by 0.7\%. Considering IoU is a stricter metric than DSC, we believe the 2.6-percent improvement is sufficient enough to demonstrate the strengths of UNet-2022 over BAT and nnUNet.

\subsection{Ablation studies of modules and strategies}
In this section, we first investigate the impact of the proposed PI block by comparing it against Swin Transformer (ST) block and ConvNeXt (CNX) block. Then, we further study the influences of the self-attention and DwConv layers in the PI block. Next, we demonstrate the performance gains of ImageNet-based pre-training over training from scratch. Finally, we propose a novel post-processing strategy, which we found empirically provides stable task performance improvements.\\

\noindent \textbf{Impact of the PI block.} Table~\ref{main_block} presents the comparisons of different building blocks, including blocks used in Swin Transformer~\cite{liu2021swin}, ConvNeXt~\cite{liu2022convnet}, and our UNet-2022. We see that the CNX block performs slightly better than the ST block in average. Nonetheless, our PI block obviously surpasses the CNX block by 1.5\% in the average DSC while largely improving HD95 by approximate 5mm. The underlying reason why our PI block performs better than the ST and CNX blocks is that the latter two blocks only use isomorphic operations, i.e., self-attention in the ST block and DwConv in the CNX block. This characteristic makes them lack the ability to capture dynamics across different dimensions.\\

\begin{table}[!t]

    \begin{center}
	\resizebox{0.27\textwidth}{!}{
	\begin{tabular}{|l|cc|} 
	\hline
	    \multirow{2}{*}{Methods} & \multicolumn{2}{c|}{Average}  \\ \cline{2-3}
	    & DSC $\uparrow$ & HD95 $\downarrow$ \\
		\hline
		\hline
		ST~\cite{liu2021swin} &84.42 &20.74 \\
        CNX~\cite{liu2022convnet} &84.96 &16.60 \\
        Our PI &{86.46} &{11.34} \\
        \hline
	\end{tabular}
	}
	\end{center}
\caption{Comparisons of major building blocks used in Swin Transformer (ST), ConvNeXt (CNX), and UNet-2022 (PI).}
\label{main_block}
\end{table}

\begin{table}[!t]
    \begin{center}
	\resizebox{0.27\textwidth}{!}{
	\begin{tabular}{|l|cc|} 
	\hline
	    \multirow{2}{*}{Methods} & \multicolumn{2}{c|}{Average}  \\ \cline{2-3}
	    & DSC $\uparrow$ & HD95 $\downarrow$  \\
		\hline
		\hline
		UNet-2022 &86.46 &11.34 \\
        \hline
        - SA &85.20 & 14.96 \\
        \hline
	\end{tabular}
	}
	\end{center}
\caption{Investigation of self-attention (SA) and DwConv layers in the PI block.}
\label{abla_sy}
\end{table}
\begin{table}[!t]
    \begin{center}

	\resizebox{0.27\textwidth}{!}{
	\begin{tabular}{|l|cc|} 
	\hline
	    \multirow{2}{*}{Methods} & \multicolumn{2}{c|}{Average}  \\ \cline{2-3}
	    & DSC $\uparrow$ & HD95 $\downarrow$  \\
		\hline
		\hline
        UNet-2022  & 86.46 &11.34 \\
        - Pre-training &84.87 &15.04 \\
        \hline
        ConvNeXt &84.96 &16.60 \\
        - Pre-training &84.32 &18.02 \\
        \hline
	\end{tabular}
	}
    \end{center}
\caption{Impact of the ImageNet-based pre-training. Specifically, we replace the default encoder in UNet-2022 with ConvNeXt to demonstrate the superiority of proposed modules.}
\label{pretrain}
\end{table}

\begin{table}[!t]
    \begin{center}
	\resizebox{0.27\textwidth}{!}{
	\begin{tabular}{|c|cc|} 
	\hline
	    \multirow{2}{*}{Step size} & \multicolumn{2}{c|}{Average}   \\ \cline{2-3}
	    & DSC $\uparrow$ & HD95 $\downarrow$\\
		\hline
		\hline
		\textsc{0.5$\times$crop size} &86.27 &13.95 \\
        \textsc{0.2$\times$crop size} &{86.46} &{11.34} \\
        \hline
	\end{tabular}
	}
	\end{center}
\caption{Impact of the inference step size on testing performance.}
\label{step}
\end{table}

\noindent \textbf{Influences of dynamics across the space and channels.} In Table~\ref{abla_sy}, we investigate the impacts of dynamics across dimensions. Firstly, we remove the self-attention layer from the PI block. As a result, the resulting building blocks fail to explore dynamics across different spatial positions, as they only contain convolution operations. This failure can be verified by the observable task performance drop in the second row of Table~\ref{abla_sy}. \\

\noindent \textbf{Impact of ImageNet-based pre-training.} We use ImageNet-based pre-training to boost the segmentation performance. We replace the default encoder in UNet-2022 with the recently proposed ConvNeXt~\cite{liu2022convnet}, and compare the modified ConvNeXt-based UNet-2022 with our proposed version on Synapse. As shown in Table~\ref{pretrain}, we see that ImageNet-based pre-training plays a vital role in both UNets, providing observable performance gains over training from scratch. More importantly, we find that ImageNet-based pre-training brings more benefits to our UNet-2022 instead of the ConvNeXt-based version. For instance, introducing pre-training boosts the average DSC by 1.5\% and improves HD95 by nearly 5mm. \\

\noindent \textbf{Impact of the inference step size.} As aforementioned, we adjust the inference step size when sampling sliding windows on Synapse. Here, we present the impact of the step size in Table~\ref{step}. When we set the step size to \textsc{0.2$\times$crop size}, we find that it performs better than \textsc{0.5$\times$crop size}, bringing about 2.6mm improvement in HD95. This comparison is easy to understand as smaller step sizes result in more sliding windows, which helps the segmentation model to aggregate more local information.\\ 

\noindent \textbf{Visualization results.} Due to the limited space of the main text, we move the visualization results of all 4 tasks to the supplementary material.

\section{Conclusion}
\label{sec:conclusion}
In this paper, we analyze the advantages of self-attention and convolution from a perspective of weight allocation. We show that self-attention and convolution assign dynamic weights to different dimensions in the feature space. On top of this analysis, we propose to build a non-isomorphic block by simply parallelizing self-attention and convolution operations. The resulting UNet-2022 achieves quite competitive performance in a range of medical image segmentation tasks, sometimes outperforming nnUNet by nearly 4\%. In the future, we will investigate how to appropriately incorporate self-supervised learning~\cite{zhou2017sunrise,zhou2021models,zhou2021preservational,zhou2022generalized,yu2020difficulty} into UNet-2022 since we found pre-training plays a vital role in advancing the segmentation performance.


{\small
\bibliographystyle{ieee_fullname}
\bibliography{egbib}
}

\end{document}